\documentclass[12pt]{article}
\usepackage{graphicx}
\begin{document}
\centerline{\Large \bf Universality of the Threshold for Complete Consensus}
\centerline{\Large \bf for the Opinion Dynamics of Deffuant et al.}

\bigskip
\bigskip

\centerline{Santo Fortunato}

\vskip0.3cm

\centerline{Fakult\"at f\"ur Physik, Universit\"at Bielefeld, 
D-33501 Bielefeld, Germany}

\noindent

\centerline{{\tt e-mail: fortunat@physik.uni-bielefeld.de}}

\bigskip

\begin{abstract}

In the compromise model of Deffuant et al., 
opinions are real numbers between $0$ and $1$ and two agents
are compatible if the difference of their opinions is smaller 
than the confidence bound parameter $\epsilon$.
The opinions of a randomly chosen pair of compatible agents 
get closer to each other.
We provide strong numerical evidence that the threshold value of 
$\epsilon$
above which all agents share the same opinion in the final configuration is 
$1/2$, independently of the underlying social topology.

\end{abstract}

\bigskip


\bigskip

The last few years witnessed many attempts to describe society as a physical
system \cite{weidlich}, with people playing the role of atoms or classical spins undergoing 
elementary interactions. There are meanwhile several models to explain
how hierarchies \cite{bonabeau} and consensus \cite{Axel}-\cite{staufrev} 
may originate in a society.

In this paper we focus on a consensus model, that of Deffuant et al.
\cite{Deff}. It is a model with binary interactions, i.e. where 
the opinions of the agents are modified pairwise, according to a 
compromise strategy. One starts from a graph with $N$ vertices, which 
are the agents of the society. The edges of the graph represent the 
relationships between the agents and
interactions can take place only between
neighbouring agents. 
Next, opinions are randomly distributed among the agents;
the opinions, usually indicated with $s$, are real numbers in the range $[0,1]$. 
Furthermore two parameters are introduced, the confidence bound $\epsilon$
and the convergence parameter $\mu$. They are both real numbers, 
which take values in $[0,1]$ and $[0,1/2]$, respectively.
The dynamics is very simple: one chooses 
a pair $\{i,j\}$ of neighbouring agents and checks whether 
$|s_i-s_j|<\epsilon$. If this is not true we do nothing.
Otherwise the agents get the new opinions
${s^{\prime}}_i=s_i+\mu\,(s_j-s_i)$ and ${s^{\prime}}_j=s_j+\mu\,(s_i-s_j)$.
This means that the opinions of the agents shift towards 
each other, by a relative amount $\mu$.
In the particular case $\mu=1/2$, the 
two opinions jump to their mean.
By repeating the procedure one sees that the agents 
group in opinion clusters until, at some stage, a stable configuration
is reached. Stable configurations can only be superpositions of
$\delta$-functions in the opinion space, such that 
the opinion of any $\delta$ is 
farther than $\epsilon$ from all others.
In this case, in fact,
if we take a pair of agents, their opinions are either identical, because the
agents
belong to the same cluster, or they differ more than $\epsilon$: in both cases 
nothing happens.
The number of clusters in the final configuration depends on the value
of $\epsilon$. In particular, there is a special $\epsilon_c$ such that, for 
$\epsilon>\epsilon_c$, all agents belong to a single cluster.
We will show here that $\epsilon_c=1/2$, no matter what kind of graph 
one takes to represent the relationships among the agents.
We study the question
numerically by means of Monte Carlo simulations. We analyzed four different 
graph structures: 

\begin{itemize}
\item{a complete graph, where everybody talks to everybody \cite{Deff};}
\item{a square lattice;}
\item{a random graph a la Erd{\"o}s and R{\'e}nyi \cite{erdos};}
\item{a scale free
graph a la Barab{\'a}si-Albert \cite{BA}.}
\end{itemize}
 
We update the opinions of the agents
in ordered sweeps over the population.  For any agent we randomly select
one of its neighbours as partner of the interaction. 
The program stops if no agent changed opinion after an iteration; since 
opinions are double precision real numbers, our criterion is to check whether any
opinion varied by less than $10^{-9}$ after a sweep.
The results do not depend on the value of the convergence parameter $\mu$,
so we have always set $\mu=1/2$.
Our method is quite simple. For a given population $N$ and 
confidence bound $\epsilon$ we collected a number of samples 
ranging from $500$ to $1000$. Once the system has reached its final 
configuration, we check whether all agents belong to the same cluster or not.
The fraction of samples with a single final opinion cluster is the 
probability $P_c$ to have complete consensus, that we study as a function of
$\epsilon$.

\begin{figure}[hbt]
\begin{center}
\includegraphics[angle=-90,scale=0.5]{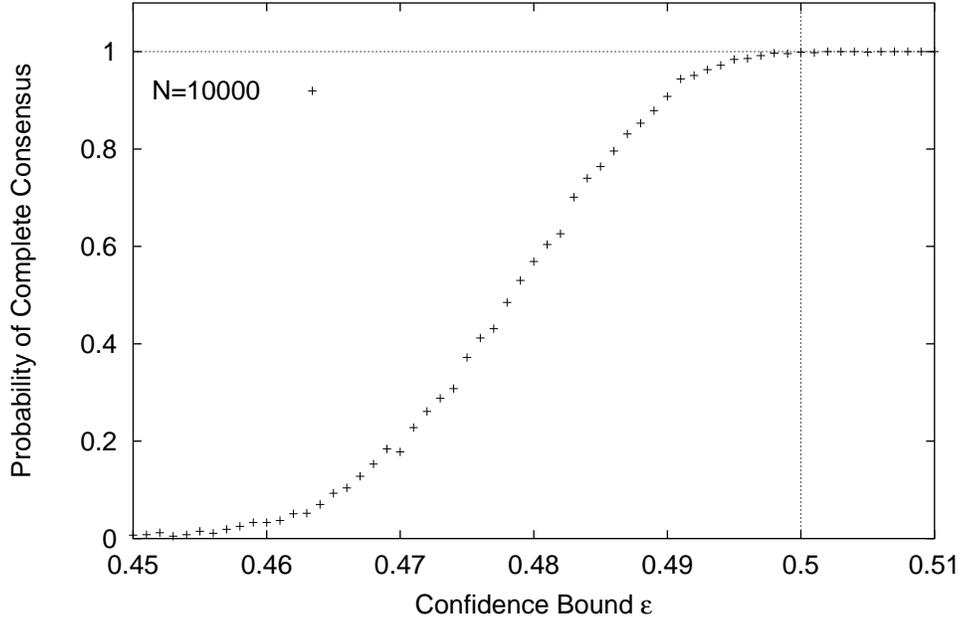}
\end{center}
\caption{\label{fig1}Fraction of samples with a single cluster of opinions 
in the final configuration, for a society where everybody talks to everybody.}
\end{figure}
\begin{figure}[hbt]
\begin{center}
\includegraphics[angle=-90,scale=0.5]{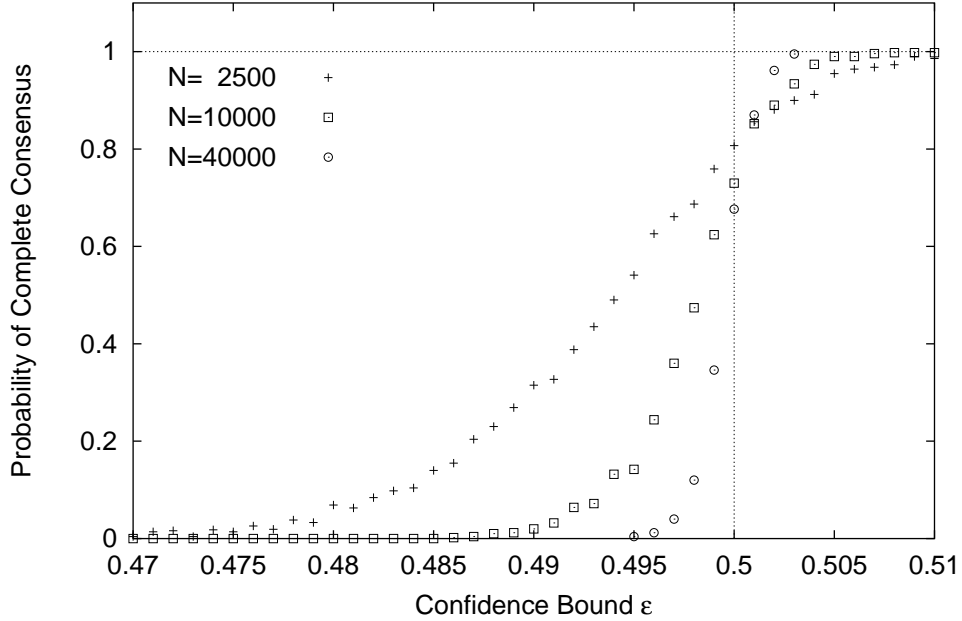}
\end{center}
\caption{\label{fig2}Fraction of samples with a single cluster of opinions 
in the final configuration, for three different populations. The 
agents sit on the sites of a square lattice, with periodic boundary conditions.}
\end{figure}
\begin{figure}[hbt]
\begin{center}
\includegraphics[angle=-90,scale=0.5]{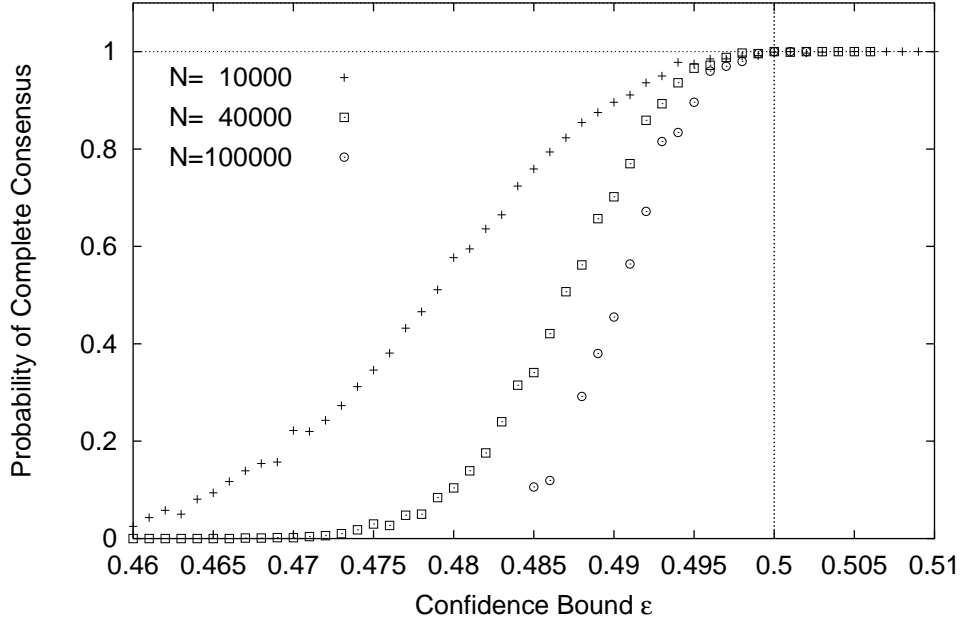}
\end{center}
\caption{\label{fig3}As Fig. \ref{fig2}, but for a random graph a la Erd{\"o}s
  and R{\'e}nyi.} 
\end{figure}
\begin{figure}[hbt]
\begin{center}
\includegraphics[angle=-90,scale=0.5]{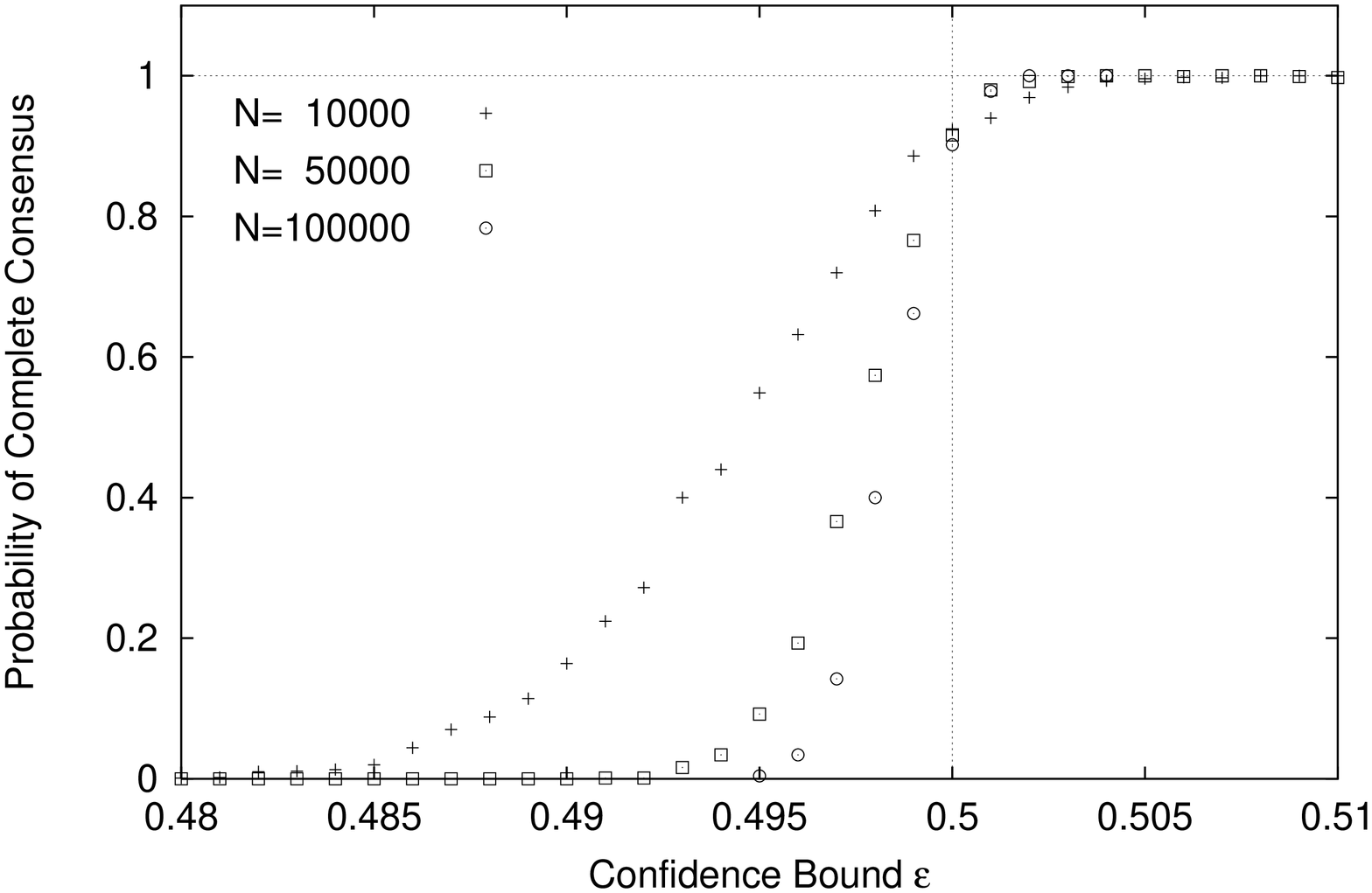}
\end{center}
\caption{\label{fig4}As Fig. \ref{fig3}, but for a scale free network a la
  Barab{\'a}si and Albert.}
\end{figure}

For a society where all agents interact with each other, a beautiful analysis 
was carried on by Ben-Naim et al. \cite{bennaim}.
In contrast to us they fixed the confidence bound to $1$ and introduced a
parameter $\Delta$ to mark the bounds of the opinion space, which goes from
$-\Delta$ to 
$\Delta$, but the results can be 
simply translated in our conventions. They derived a rate equation 
for the dynamics and solved it numerically,
finding that for $\Delta>1$, which
corresponds to $\epsilon>1/2$, all agents share the same opinion. 
Because of that, we did not study this case in detail, but we performed 
some simple simulations to test our method. Fig. \ref{fig1} shows the behaviour
of $P_c$ with $\epsilon$ for a society with $10000$ agents. 
We observe that $P_c$ is basically zero up to $\epsilon\,\sim\,0.46$, 
it rises rapidly for $0.46<\epsilon<1/2$ and 
it saturates to one for $\epsilon>1/2$, which is compatible
with the result of \cite{bennaim}.
Fig. \ref{fig1} suggests that $P_c$ may converge to a step function $\theta$
in the limit $N\rightarrow\infty$. This can be best verified by 
using several values of $N$ and analyzing what happens when $N$ increases, and this
is the strategy we adopted for the other social topologies. 

Let us now examine the situation for an ordered structure like a lattice.
We took a square lattice with periodic boundary conditions, so each agent has exactly
four neighbours. Early studies of the Deffuant dynamics on a lattice
were carried on in \cite{weisbuch}.
In Fig. \ref{fig2} we again plot $P_c$ as a function 
of $\epsilon$, but this time we repeated the procedure for three different
population sizes, $N=2500$, $10000$ and $40000$. The convergence to a step
function with threshold $1/2$ is manifest.  

In a realistic model of a society, it is neither true that everybody knows
everybody else
(unless one considers small communities), nor that every person has the same
number of friends. A plausible model is given by a random graph (or random network).
We considered two types of random graphs, the classical model of 
Erd{\"o}s and R{\'e}nyi \cite{erdos} and the scale free model proposed by 
Barab{\'a}si and Albert \cite{BA}, which has attracted an exceptional 
attention in the last years \cite{netw}.

The random graph of Erd{\"o}s and R{\'e}nyi is characterized by a parameter
$p$, which is the connection probability of the nodes. One assumes that 
each of the $N$ nodes of the graph has probability
$p$ to be linked to any other node.
In this way, the total number of edges $m$ of the graph is 
$m=pN(N-1)/2$ and the average {\it degree} of the graph, i.e. the 
average number of neighbours of a node, is $k=p(N-1)$ which 
can be well approximated by $pN$ when $N\rightarrow\infty$.
In order to have a finite degree, the product $pN$ must then be finite.
In our simulations we built graphs with the same average degree $k=pN=400$, and
number of nodes $N=10000$, $40000$ and $100000$.
Fig. \ref{fig3} shows the behaviour of $P_c$ with $\epsilon$ 
for this special topology: $P_c$ equals one for $\epsilon>1/2$. 
The fact that for any $\epsilon<1/2$ $P_c$ decreases with $N$
confirms the impression that, in the limit $N\rightarrow\infty$,
$P_c=0$ for $\epsilon<1/2$.

Finally, we examined our problem for agents sitting 
on the nodes of a scale free network.
This topology was adopted for the Deffuant model 
in \cite{sta1} and \cite{sta2} as well.
To build the network we must 
specify the outdegree $m$ of the nodes, i.e. the number of edges which originate
from a node. The procedure is dynamic; one starts from $m$ nodes which are all
connected to each other and adds further $N-m$ nodes one at a time.
When a new node is added, it selects $m$ of the preexisting nodes as neighbours,
so that 
the probability to get linked to a node is proportional
to the number of its neighbours. In all networks created in this way
the number of agents with degree $k$ is proportional to $1/k^3$ for $k$ large,
independently of $m$; here we chose $m=3$.
The results are illustrated in Fig. \ref{fig4}. Once more, we get the 
same pattern observed in the previous cases.

We have then discovered a general feature of the opinion dynamics 
of Deffuant et al.: no matter how society is structured, for $\epsilon>1/2$
there is always complete consensus, whereas for $\epsilon<1/2$ there are 
at least two different opinion clusters. When we are slightly below $1/2$ 
one cluster includes all
agents except just a few (or even a single agent!). 
There is a simple argument to convince ourselves 
why this happens. If $\epsilon>1/2$ there cannot be more than one large opinion
cluster. As a matter of fact, because of the symmetry of the model, the 
disposition of the clusters in the opinion space is also symmetric
with respect to the center opinion $s=1/2$. 
Let us concentrate on the two
clusters which lie close to the extremes of the opinion interval.
The agents whose opinions lie close to $0$, for instance, can interact with
agents with opinions in the range $[0,\epsilon]$. Due to the symmetry 
of the dynamics, the center $\epsilon_l$ of the 
peak will approximately coincide with the middle point 
of the latter range, so that $\epsilon_l\,\sim\,\epsilon/2$.
Analogously, the center of the rightmost peak $\epsilon_r\,\sim\,1-\epsilon/2$.
In this way, the 
distance between the rightmost and the leftmost cluster 
$\epsilon_r-\epsilon_l\,\sim\,1-\epsilon$, which 
is smaller than $\epsilon$ if $\epsilon>1/2$.
The two clusters will then fuse at some stage to a unique cluster at $s=1/2$.
The agents of this big central cluster are compatible with all remaining
agents, as the maximal difference between $1/2$ and
all possible opinions is $1/2<\epsilon$.
At the end of the day, all agents which are not already in the major cluster will be 
sooner or later attracted by it. The argument we exposed is obviously
independent of the way the agents are connected to each other, but it only 
relies on the dynamics. However, it is not
a rigorous proof, and this is the reason why we recurred to the numerical
simulations we have presented. 

We believe that the
result holds as well 
for other versions of the Deffuant model. For example, one could 
assign different values of $\mu$ to all agents, like in \cite{tesi}.
As long as the $\mu$'s are distributed
independently of the opinions of the agents
the result should still be valid, for any distribution. We performed some tests,
by using a uniform and an exponential distribution for $\mu$, and they 
confirm our expectation.

\bigskip

I am indebted to D. Stauffer for introducing me into this fascinating field and
for many suggestions and comments.
I gratefully acknowledge the financial support of the DFG Forschergruppe
under grant FOR 339/2-1.

\end{document}